\newcommand{\figref}[1]{fig.\ \ref{#1}}

\newcommand{\eqnref}[1]{eq.\ (\ref{#1})}
\newcommand{\Eqnref}[1]{Eq. (\ref{#1})}

\newcommand{\eqnsref}[2]{eqs.\ (\ref{#1}) and (\ref{#2})}
\newcommand{\eqndref}[2]{eqs.\ (\ref{#1})--(\ref{#2})}
\newcommand{\eqnsrefthree}[3]{eqs.\ (\ref{#1}), (\ref{#2}) and (\ref{#3})}

\newcommand{\ie}[0]{\emph{i.e.}}
\newcommand{\eg}[0]{\emph{e.g.}}

\pdfoutput=1
\documentclass[doublecol,amsmath,amssymb]{epl2}
\usepackage{graphicx,bm,verbatim,hyperref}

\begin{document}
\bibliographystyle{eplbib} 

\title{Global disorder transition in the community structure of large-$q$ Potts systems}

\abstract{We examine a global disorder transition when identifying community 
structure in an arbitrary complex network. 
Earlier, we illustrated [\emph{Phil. Mag.} \textbf{92}, 406 (2012)] that ``community 
detection'' (CD) generally exhibits disordered (or unsolvable) and ordered 
(solvable) phases of both high and low computational complexity along with 
corresponding transitions from regular to chaotic dynamics in derived systems.
Using an exact generalized dimensional reduction inequality, multivariate Tutte 
polynomials, and other considerations, we illustrate how increasing the number 
of communities $q$ emulates increasing the heat bath temperature $T$ for a 
\emph{general} weighted Potts model, leading to global disorder in the community 
structure of \emph{arbitrary} large graphs.
Dimensional reduction bounds lead to results similar to those suggested 
by mean-field type approaches.
Large systems tend toward global insolvability in the limit of large $q$
above a crossover temperature $T_\times\approx L|J_e|/\left[N\ln{} q\right]$ 
where $|J_e|$ is a typical
interaction strength, $L$ is the number of edges, and $N$ is the number of nodes.
For practical system sizes, a solvable phase is generally accessible at low $T$.
The global nature of the disorder transition does not preclude solutions by local
CD algorithms (even those that employ global cost function parameters) as long 
as community evaluations are locally determined.}

\author{Peter Ronhovde \and Dandan Hu \and 
Zohar Nussinov\thanks{E-mail: \email{zohar@wuphys.wustl.edu}}}
\shortauthor{P. Ronhovde, D. Hu and Z. Nussinov}
\institute{Department of Physics, Washington University in St. Louis -
Campus Box 1105, 1 Brookings Drive, St. Louis, MO 63130, USA}

\maketitle{}


\section{Introduction}
Methods of statistical physics have enlarged the understanding of complex
networks \cite{ref:newmanphystoday}.
In particular, community detection (CD) \cite{ref:fortunatophysrep}
attempts to identify ``mesoscopic'' structure within these systems.
Applications of CD are extremely broad, and numerous methods have been 
leveraged to solve it \cite{ref:clausetlarge,ref:smcd,
ref:blondel,ref:rosvallmaprw,ref:gudkov,ref:kumpulacliqueperc,ref:lanc,
ref:rzmultires,ref:barberLPA,ref:lancLFRcompare,ref:chengshen,ref:shenchengspectral}.
The problem complexity and related aspects of community ``detectability'' 
were studied for an ``absolute Potts model'' (APM) \cite{ref:rzlocal,ref:huCDPTsgd}, 
modularity \cite{ref:goodMC,ref:nadakuditiSBM}, 
and mean-field type (cavity) approaches \cite{ref:reichardtstruct,ref:decelleKMZPT}
where the latter references \cite{ref:nadakuditiSBM,ref:reichardtstruct,ref:decelleKMZPT}
examined general cluster detectability transitions in a special class of stochastic 
block models.
Ref.\ \cite{ref:dorogovtsevRMP} reviewed critical phenomena in complex networks.

We illustrated \cite{ref:huCDPTsgd} that the APM, along with other CD approaches, 
exhibits solvable phases of ``easy'' or ``hard'' complexity and unsolvable phases 
with spin-glass-type or other transitions coinciding with transitions from ergodic 
to non-ergodic dynamics in mechanical analogs.  
Bashan \etal{} \cite{ref:bashanphystrans} showed that transitions in network 
topology have physiological significance, implying that network transitions 
have relevance beyond detectability/solvability thresholds.
We also previously demonstrated \cite{ref:HRNimages} how distinct phases 
of the CD problem affect image segmentation applications.
Other authors covered disorder transitions for random-bond Potts models
\cite{ref:juhaszrbPMlargeq,ref:mercaldorbPMlargeq} and zeros of the partition
function \cite{ref:changlargeqzeroes} in the limit of a large number of Potts 
spin flavors ($q \gg 1$).

\begin{figure}
\centering
\includegraphics[width=0.6\columnwidth]{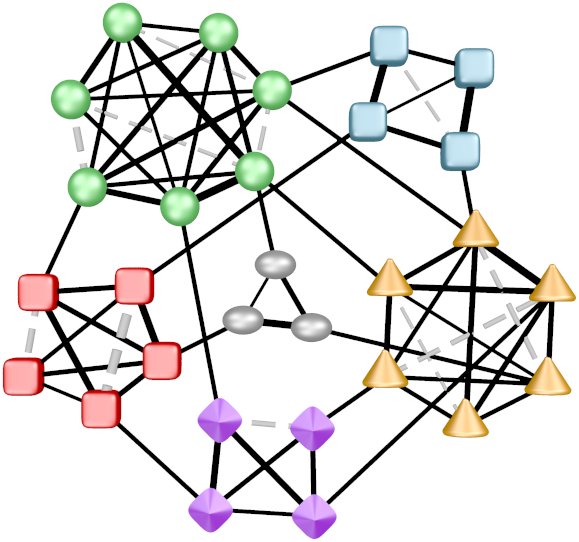}
\caption{(Color online) The schematic illustrates a community partition
(distinct node shapes and colors) showing relevant structure in a graph
with ferromagnetic (solid, black lines) and antiferromagnetic interactions
(gray, dashed lines).
Line thickness indicates the relative interaction strength $J_e$.
Antiferromagnetic (``adversarial'' with $J_e<0$) and non-interacting 
(``neutral,'' unconnected with $J_e=0$) relations break up well-defined 
communities.}
\label{fig:communities}
\end{figure}

As depicted in \figref{fig:communities}, CD attempts to partition a graph into 
$q$ optimally disjoint subgraphs (or communities). 
Optimal values of $q$ may be determined via multiscale methods
\cite{ref:arenasmultires,ref:lanc,ref:kumpulamultires,ref:rzmultires,ref:rosvallmultires}.
Ref.\ \cite{ref:leskoveclargenetworks} discussed the absence of large clusters 
in large real networks (\ie{}, $q$ is large).

We investigate a \emph{general} weighted Potts model 
on an \emph{arbitrary} graph with $q \gg 1$, and we illustrate how 
increasing $q$ emulates increasing the temperature $T$.
For CD, it implies that whenever an algorithm or cost function can be 
represented as a weighted Potts model, then \emph{large systems are 
inherently disordered} on a global level above a crossover temperature 
$T_\times$.
The result encompasses a wide variety of CD methods including optimizing 
modularity \cite{ref:gn},
a Potts model applying a ``configuration null model'' (CMPM) \cite{ref:smcd,ref:traagPRE},
an Erd{\H o}s-R{\' e}nyi Potts model \cite{ref:reichardt,ref:smcd},
a ``constant Potts model'' \cite{ref:traaglocalscope},
``label propagation'' \cite{ref:LPA,ref:barberLPA},
the APM \cite{ref:rzmultires,ref:rzlocal},
and others \cite{ref:blatt,ref:ispolatov,ref:hastings}.
We speculate that the disorder persists for systems with external magnetic
interactions \cite{ref:ellismonaghanexternalH} as well as directed and 
multipartite graphs.

While the result applies to general Potts models on arbitrary large
graphs, it only  implies disorder on a global, as opposed to local, level
for large-$q$ networks with bounded coordination numbers and vertex count.
Refs.\ \cite{ref:rzlocal,ref:traaglocalscope} were shown to avoid a
``resolution limit'' imposed by global \emph{cost function} parameters
on some models
\cite{ref:gn,ref:fortunato,ref:smcd,ref:kumpulaResLim,ref:lancfortunatomod},
but all weighted (or unweighted) Potts models would be subject
to the \emph{global} disorder imposed by large $q$.
This global disorder can be mitigated or avoided by solving the system 
locally or at sufficiently low $T$, but the latter condition exists 
in competition with beneficial thermal annealing effects of increased 
temperature at sufficiently low $T$ (``order by disorder'') 
\cite{ref:huCDPTsgd,ref:huCDPTlong}.

The disorder induced by large $q$ is quantitatively different from that
caused by high noise (extraneous intercommunity edges) in a network
\cite{ref:rzlocal,ref:huCDPTsgd,ref:huCDPTlong}.
A ``glassy'' transition due to noise may persist as $T\to 0$ because the 
solution algorithm is frustrated by a complex energy landscape exhibiting 
numerous local minima.

\section{Potts Hamiltonian} \label{sec:potts}
We consider a general spin-glass-type Potts model Hamiltonian
\begin{equation}
  H(\{\sigma\}) = - \sum_{i\neq j} J_{ij} \delta(\sigma_i,\sigma_j),
  \label{eq:Jijpotts}
\end{equation}
where $J_{ij}$ is the interaction strength between spins $i$ and $j$
and $\delta(\sigma_i,\sigma_j)=1$ if $\sigma_i=\sigma_j$ and $0$ otherwise.
For CD, it is convenient separate the ferromagnetic ($J_{ij}>0)$ and 
antiferromagnetic ($J_{ij}\le 0$) interactions 
\begin{equation}
  H(\{\sigma\}) = - \sum_{i\neq j}\big[ w_{ij} A_{ij}
     - u_{ij}\left(1-A_{ij}\right)\big] \delta(\sigma_i,\sigma_j).
  \label{eq:generalpotts}
\end{equation}
Given $N$ nodes, 
$\{A_{ij}\}$ is the adjacency matrix where $A_{ij}=1$ if nodes $i$ and $j$ 
are connected by a ferromagnetic edge and is $0$ otherwise.
$w_{ij}>0$ and $u_{ij}\ge 0$ are ferromagnetic and antiferromagnetic 
edge weights, respectively.
Each spin $\sigma_{i}$ may assume integer values in the range
$1\leq\sigma_{i}\leq q$ where $q$ is usually dynamically determined.
Node $i$ is a member of community $k$ when $\sigma_{i}=k$.

The antiferromagnetic weights $u_{ij}$ provide a ``penalty function'' 
enabling a non-trivial CD ground state for an arbitrary graph.
Some models incorporate a weight factor, generally on the $u_{ij}$ 
term, which allows the model to span different network scales in qualitatively 
similar ways.
The APM penalizes ``neutral'' relationships (\ie{}, generally 
$u_{ij}\equiv J_{ij} - 1$ for $J_{ij}\le 0$). 
Another model \cite{ref:traaglocalscope} incorporates weighted antiferromagnetic 
interactions into $w_{ij}$ and applies a separate penalty term.
Algorithms for modularity and the CMPM are effectively implemented with dynamic 
edge weights on the $u_{ij}$ term, but the fluctuations would be small in general 
approaching the ground state.
Local CD models for energy calculations were suggested 
in \cite{ref:fortunato,ref:kumpulaResLim}, further advocated in \cite{ref:rzlocal},
and explored in more detail in \cite{ref:traaglocalscope}.

\section{Dimensional reduction bound} \label{sec:Dbound}
We first provide rigorous bounds on the disorder transition for community 
structure using dimensional reduction inequalities 
\cite{ref:batista_nussinov_Dreduc,ref:nussinovholographies}.
In the current context, these simple, yet exact, inequalities relate a system 
in any dimension to a local ($D=0$ dimensional) system composed of a single 
vertex (or a finite collection of vertices) and its (their) neighbors. 
The derived bound has a form similar to that suggested by mean field 
considerations. 

In the thermodynamic limit, a bona fide transition may occur that marks symmetry 
breaking wherein, for infinitesimally weak applied fields that favor a particular 
state, the probability that a given spin belongs to one of the $q$ communities 
differs from $1/q$. 
From a practical standpoint, we are interested in the probability that a particular 
spin $\sigma_0$ takes on a specific ``correct'' spin value $\bar{\sigma}$ that it 
does in a low energy configuration, effectively searching for a ``planted'' 
solution $\bar{\sigma}$.

We derive upper bounds on the temperature for which the spin $\sigma_{0}$ 
attains its correct value with high confidence. 
Towards this end, we first detail general inequalities and then turn to their 
application in our case. 
We consider a partition of all spins into those of a local set $\eta$ (\ie{}, 
$\sigma_{0}$ in the single-spin case) and all other remaining spins $\psi$ 
in the system. 
The trace over all spins becomes 
$\mbox{Tr}_{\{\sigma\}} = \mbox{Tr}_{\{\psi\}}\mbox{Tr}_{\{\eta\}}$ 
and the Hamiltonian (with or without any weak applied fields) becomes a function 
involving both sets of spins $H(\{\sigma\})= H(\{\psi\}, \{\eta\})$. 
Any thermal average $\langle f(\eta) \rangle$ can be written as
\begin{equation}
\langle f \rangle = \frac{ \mbox{Tr}_{\{\psi\}} z_{\psi}  \left[\frac {\mbox{Tr}_{\{\eta\}} f(\eta_{\mathbf{i}})  
                  e^{-\beta H(\psi,\eta)}}{z_{{\{ \psi \}}}} \right] }
                {\mbox{Tr}_{\{\psi\}} z_{\psi} }.
     \label{mast}
\end{equation}
where we inserted $z_{\psi} \equiv \mbox{Tr}_{\{\eta\}} e^{-\beta H(\psi,\eta)}  \ge 0$
twice in the numerator which is valid for any Hamiltonian.
As can be readily seen, the exact $\langle f \rangle$ over the large system 
can be written as a weighted sum (with positive normalized weights, 
$p_{\psi} = z_{\psi}/ \mbox{Tr}_{ \{\psi'\} } z_{\psi' }$, that sum to unity) 
of local averages 
\begin{eqnarray}
  \langle f \rangle_{\psi} \equiv   \frac{
  \mbox{Tr}_{\{\eta\}} f(\eta) e^{-\beta H(\psi,\eta)} } { z_{\psi} }.
  \label{final}
\end{eqnarray}
Thus, $\langle f \rangle \leq \langle f \rangle_{\overline{\psi}}$,
where $\overline{\psi}$ is a particular set of the spins $\psi$ that
maximizes $\langle f \rangle_{\psi}$. When we substitute $\psi =
\overline{\psi}$ in $H$, we obtain a local Hamiltonian
$H(\overline{\psi}, \eta)$ in the spins $\eta$.

If we set $f\equiv\delta\left({\sigma_0,\bar{\sigma}}\right)$, then 
the mean value of $\langle f\rangle$ will correspond to the probability 
$\langle f\rangle = P(\sigma_0=\bar{\sigma})$.
When computing the internal general trace over $\eta$, we evaluate 
$\langle f \rangle_{\overline{\psi}}$, in the case of a single spin 
at the origin $\sigma_0$ averaged over its $q$ possible states.  
Applying $\langle f \rangle \leq \langle f \rangle_{\overline{\psi}}$ to $\langle \delta\left({\sigma_0,\bar{\sigma}}\right)\rangle$, 
we obtain a generous upper bound. 
If the interaction between $\sigma_{j}$ at site $j$ 
and $\sigma_0$ at the origin is larger than $0$ (\ie{}, $J_{j0}>0$), 
then $\bar{\psi_j}=\bar{\sigma}$.
If $J_{j0}<0$, then we set $\bar{\psi_j}\neq \bar{\sigma}$. 
With this set of $\psi$ values, 
\begin{equation}
  p =   \langle f\rangle
    \le \frac{e^{\beta \bar{J_0}}}{e^{\beta \bar{J_0}} + (q-1)},
\end{equation}
where $\bar{J_0}=\frac{1}{2}\sum_j J_{j0}\left[ 1+\mbox{sgn}(J_{j0}) \right]$.

On practical benchmarks, we are interested in cases where $p$ 
exceeds some threshold value $p^*$.
The inverse temperature $\beta^*$ at which the 
probability exceeds $p^*$ is 
\begin{equation}
  \beta^* = \frac{1}{\bar{J_0}}\ln{} \left[ \frac{p^*(q-1)}{(1-p^*)} \right].
\end{equation}
At high $q$ with $p^{*}=1/2$, this leads to rigorous upper bound (UB) 
for the associated crossover temperature 
\begin{equation}
  T_\times^\mathrm{UB} = \frac{\bar{J_0}}{k_B \ln{} q}.
  \label{eq:TcrossUB}
\end{equation}
For $T>T_\times^\mathrm{UB}$, the correct assignment
can only be determined with a probability $p^* = 1/2$.
If the exchange constants $J_{j0}$ and the coordination number of $\sigma_0$
are finite and do not match or exceed the $\ln{} q$ dependence in the denominator, 
the system is unsolvable at any positive temperature as $q \to \infty$. 
The bound of \eqnref{eq:TcrossUB} for the node at the origin is local, 
so it may change for other nodes. 

In practice, some parts of the network can exhibit structure at higher 
temperatures which serves as a bottleneck for global ordering.  
Generally, the bounds of 
$\langle f \rangle \leq \langle f \rangle_{\overline{\psi}}$ 
enable a reduction of the full physical system to a related problem that 
occupies a reduced $D$-dimensional sub-volume of the entire system. 
If we define the external state $\psi$ as a set of spins with the average 
spin value, then the resulting average becomes a mean-field average.

We discuss a general representation for Potts model where, 
with it and related approaches, we estimate the form of the cross-over 
(or transition) temperature $T_\times$ from a viable low temperature 
ordered phase to a high temperature disordered regime. 
The scaling in these results is similar to that of the rigorous bound 
in \eqnref{eq:TcrossUB}.

\section{Multivariate Tutte polynomial estimate} \label{sec:MVTP}
The multivariate Tutte polynomial \cite{ref:jacksonmvtutte} is defined
as a subgraph expansion over $\mathcal{A}\subseteq \mathcal{E}$
of a graph $G = (V,\mathcal{E})$ where $V$ and $\mathcal{E}$ are the sets
of vertices and (ferromagnetic and antiferromagnetic) edges, respectively.
\begin{equation}
  Z(G;q,\mathbf{v}) = \sum_{\mathcal{A} \subseteq \mathcal{E}}
       q^{k(\mathcal{A})} \prod_{e' \subseteq \mathcal{A}} v_{e'}
  \label{eq:ZarbitraryGq}
\end{equation}
where $k(\mathcal{A})$ is the number of connected components
of $G_A = (V,\mathcal{A})$, $v_e = \exp{} (\beta J_e) -1$, 
and $J_e$ is the interaction strength of edge $e$.
In CD, large $q$ necessarily implies a large number of nodes $|V|=N$.

For two disjoint partitions $A$ and $B$ with $G=A\cup B$,
$Z(G;q,\mathbf{v}) = Z(A;q,\mathbf{v}_A) Z(B;q,\mathbf{v}_B)$ where
$\mathbf{v}_A$ and $\mathbf{v}_B$ are the edge weights in the respective
subgraphs.
For unweighted systems, the interaction strength is $J_e\equiv J=\pm 1$
where $+$ and $-$ correspond to ferromagnetic or antiferromagnetic
interactions, respectively.

There is a slight terminology distinction between CD and the energy
contributions in \eqnref{eq:Jijpotts}.
Edges with $J_e>0$ correspond to the $w_{ij}$ ferromagnetic (``friendly''
or ``cooperative'') interactions in \eqnref{eq:generalpotts}, and $J_e\le 0$
relates to the $u_{ij}$ anti-ferromagnetic (neutral and ``adversarial''
in some models) or absent (neutral in some models) interactions.
The edge effect is conceptually consistent with CD for $J_e>0$, but 
antiferromagnetic weights are also related by an edge when calculating 
\eqnref{eq:ZarbitraryGq}.
That is, an interaction exists, but it is antiferromagnetic in nature.
In CD, repulsive antiferromagnetic interactions correspond to adversarial 
relationships which act like neutral (unconnected) relations that 
hinder community structure.

For large $T$, we require $T\gg \max_{e\in\mathcal{E}} |J_{e}|$.
The leading order terms for an \emph{arbitrary} graph are due to
$\mathcal{A}_\emptyset=\{\emptyset\}$ and $\mathcal{A}_e=\{e\}$
for each edge $e\in\mathcal{E}$. 
We also include the last $\mathcal{A}=\mathcal{E}$ term of $G$ 
which is addressed later in the text.
\begin{equation}
  Z(G;q,\mathbf{v}) =
          q^N \left( 1 + \sum_{e'=1}^{|\mathcal{E}|} \frac{v_{e'}}{q}
           + \cdots + q^{k(G)-N} \prod_{f'=1}^{|\mathcal{E}|} v_{f'} \right)
  \label{eq:ZarbitraryGlargeqkG}
\end{equation}
which applies to arbitrarily large systems of size $N$.
The free energy per site $f=-\frac{k_B T}{N}\ln{} Z$ becomes
\begin{equation}
  f \approx -k_B T\ln{} q - \frac{k_B T}{N} \sum_{e'=1}^{|\mathcal{E}|}
                     \frac{v_{e'}}{q}
  \label{eq:farbitraryGWsumve}
\end{equation}
where we invoke the small $x$ approximation, $\ln{} (1+x)\approx x$,
and neglect the last $\mathcal{A}=\mathcal{E}$ term 
in \eqnref{eq:ZarbitraryGlargeqkG}.
For large $T$, $v_e\approx J_e/(k_B T)$ and
\begin{equation}
  f \approx -k_B T\ln{} q - \frac{1}{N} \sum_{e'=1}^{|\mathcal{E}|}
                     \frac{J_{e'}}{q}.
  \label{eq:farbitraryGWsumT}
\end{equation}
When compared to \eqnref{eq:farbitraryGWsumq} for large $q$, 
it implies that $q$ emulates $T$ when $T$ is large.

When $q\gg \max_{e\in\mathcal{E}} |v_{e}|$, the leading order terms 
in $1/q$ are still the first two in \eqnref{eq:ZarbitraryGlargeqkG}. 
As $T\to 0$, higher order terms in $v_e$ become increasingly important
which is dominated by the last subgraph term for $\mathcal{A}=\mathcal{E}$. 
The three displayed terms are ``universal'' since they apply equally
well to all graphs (\eg{}, lattices, Cayley trees, random graphs, etc.).
That is, they only depend on the system size $N$, the number of links $L$,
and the number of connected components $k(G)$ for the full graph $G$.

All displayed terms in \eqnref{eq:ZarbitraryGlargeqkG} are identical 
for regular lattices and similar-coordination-number Bethe lattices.
Similar results are obtained for other graphs where Bethe lattice 
approximants are only identical for these terms.
The important non-universal terms in the subgraph expansion [denoted by 
ellipsis] depend on the particular graph structure.
For the leading and last terms, respectively, the logarithms of the terms 
flesh out the typical scales of the high 
temperature entropic and low temperature energetic contributions to the 
free energy.

In the large-$q$ limit, the zeroes of $Z$ for constant $v$ provide the
relevant transition temperatures in the $N \to \infty$ limit, and the free 
energy per site becomes non-analytic. 
We can estimate disorder transition temperatures by comparing the second and 
last terms, $q^{N-1}\sum_{e'} v_{e'}$ to $q^{k(G)}\prod_{f'=1} v_{f'}$, assuming 
a ``typical'' interaction strength $|J_e|$ so that we can solve the equation.
If $q$ is large then the latter term in \eqnref{eq:ZarbitraryGlargeqkG}
will compete with the last term suggesting a crossover temperature
\begin{equation}
  T_\times\approx \frac{|J_e|}{k_B\ln{}\left(q^{\left[N-k(G)\right]/L} + 1\right)}
  \label{eq:TcrossoverLNfull}
\end{equation}
under the assumptions $N\gg 1$ and $L\gg 1$. For general $\{J_e\}$, 
we may see multiple transitions spread over a range of $T$.
In the limit as $T\to 0$, \eqnref{eq:TcrossoverLNfull} becomes
\begin{equation}
  T_\times\approx \frac{L|J_e|}{k_B\left[N-k(G)\right]\ln{} q}.
  \label{eq:TcrossoverLN}
\end{equation}
If we instead compare $v_{f'}/q$ to $1$ in \eqnref{eq:ZarbitraryGlargeqkG},
the factor $L/N$ disappears, but the logarithmic behavior in $q$ remains.
\Eqnref{eq:TcrossoverLN} diverges for an arbitrarily large complete graph
[$L=N(N-1)/2$ and $N\to\infty$], and it approaches zero as $q\to\infty$.
Often in CD, the graph is (almost) completely connected (in a topological 
sense) so $N\gg k(G)$.
For \emph{sparse} graphs, $L\propto N$, so 
\begin{equation}
  T^{\mathrm{Sparse}}_\times\approx \frac{d|J_e|}{2k_B\ln{} q}
  \label{eq:Tcrossoverd}
\end{equation}
where $d$ is an average node degree.

Above $T_\times$, the large-$q$ contributions dominate, and the system
is in a disordered state, but it is globally ordered for $T\ll T_\times$.
For moderate levels of noise in a graph, larger $d$ (with a well-defined
community structure) actually \emph{increases} the crossover temperature.
This indicates that additional noise up to an insurmountable threshold 
allows the system to explore the phase space more completely when the 
community structure is solved \cite{ref:huCDPTlong}.
For degree distributions seen in CD (\eg{}, often a power law),
the corresponding crossover temperature would spread or split into
multiple values which model the distinct features of the graph.
In the limit of large $N$, the crossover(s) would become an approximate
transition point.

In order to highlight the similarity between the large-$q$ and $T$ behaviors, 
we fix $T=T'>T_\times$, define the constant 
$J^{(q)}_e\equiv k_B T' \left\{\exp{} [J_e/(k_B T')]-1\right\}$, and rewrite 
\eqnref{eq:farbitraryGWsumve} as
\begin{equation}
  f \approx -k_B T'\ln{} q - \frac{1}{N} \sum_{e'=1}^{|\mathcal{E}|}
                     \frac{J_{e'}^{(q)}}{q} .
  \label{eq:farbitraryGWsumq}
\end{equation}
Large $q$ in \eqnref{eq:farbitraryGWsumq} emulates large $T$
in \eqnref{eq:farbitraryGWsumT}.
$J_e^{(q)}$ is exponentially weighted in $\beta'$, so a non-zero (perhaps
small) region of stability is ensured except in the presence of high noise
\cite{ref:rzlocal,ref:huCDPTsgd,ref:huCDPTlong}.
Transitions between contending minima in random embedded systems \cite{ref:huCDPTsgd},
contending states with multi-scale (\eg{}, hierarchical) structures \cite{ref:rzmultires}, 
or others may occur over a range of temperatures.

\section{Mean-field and free energy estimates} \label{sec:MFandFE}
The mean-field transition temperature for lattices with a fixed coordination 
number $d$, constant exchange $J$, and arbitrary $q$ is
\cite{ref:pottslargeqmean,ref:biskuppotts} 
\begin{equation}
  T^\mathrm{MF}_c = \frac{Jd(q-2)}{2k_B (q-1)\ln{} (q-1)}\
  \label{eq:TcrossoverMFarbq}
\end{equation}
for $q\ge 3$. 
This equation yields a large-$q$ limit of 
\begin{equation}
  T_c^\mathrm{MF}\approx \frac{dJ}{2k_B \ln{} q}
  \label{eq:TcrossoverMF}
\end{equation}
in agreement with \eqnref{eq:Tcrossoverd}.
The $q\to\infty$ limit on ferromagnetic lattices asymptotically approaches 
the mean-field theory result with translationally invariant $J$ and constant 
$d$ \cite{ref:pottslargeqmean,ref:pearcelargeqmean}.
The Gibbs-Bogoliubov-Feynman inequality also allows a method 
for deriving optimal mean-field approximations in general.

We can ascertain the same asymptotic behavior in $q$ by analyzing the free 
energy per site if we flip a spin in a ground state. 
Assuming that the energy and entropy changes are uncorrelated, the energy 
change for the node flip is $\Delta U \simeq d|J_e|$ up to an undetermined 
constant factor, and the entropy change is $\Delta S\simeq\ln{} q$ yielding 
a free energy change $\Delta F \simeq d|J_e| - k_B T\ln{} q$.
The entropy contribution dominates (see \cite{ref:changlargeqzeroes} 
on general lattices) 
$\Delta F$ above a crossover estimate 
\begin{equation}
  T_\times^\mathrm{FE}\approx \frac{d|J_e|}{k_B \ln{} q}
  \label{eq:TcrossFE}
\end{equation}
which agrees well with \eqnref{eq:Tcrossoverd}.
As we alluded earlier, the logarithms of the leading and last terms in 
\eqnref{eq:ZarbitraryGlargeqkG} trivially provide the typical entropic 
and energetic contributions to the free energy at high and low temperatures.

\begin{figure}[t!]
\begin{center}
\includegraphics[width=0.7\columnwidth]{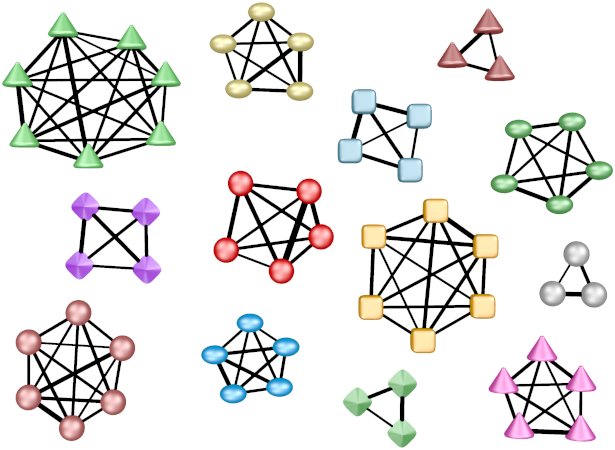}
\end{center}
\caption{(Color online) The figure depicts $q$ independent maximal sub-graphs
of size $n_i$ for $i=1$ to $q$. The line thickness illustrates the relative strengths.
With strictly ferromagnetic weights as depicted, this system is the \emph{strongest}
possible community structure since it is ``perfectly'' defined via a maximal
number of internal edges with no external edges to muddle the natural structure.
Any antiferromagnetic weights would weaken the community structure 
much like ``neutral'' relations act to break up a well-defined community.
In \eqndref{eq:TcrossUB}{eq:TcrossFE}, we show that \emph{all}
large-$q$ CD problems that are representable as a general weighted Potts model
experience \emph{global} disorder at high $q$ above a crossover temperature
$T_\times$, and we calculate the partition function and the free energy per site
of this system as a specific example in \eqndref{eq:ZcliqueslargeqT}{eq:triTcross}.}
\label{fig:cliqueexample}
\end{figure}

\section{Non-interacting cliques example} \label{sec:NICexample}
The \emph{most} strongly defined community structure is a system of $q$ 
non-interacting cliques (maximally connected sub-graphs with no intercommunity 
relationships that obscure community structure).
We define $q$ weighted cliques with sizes $n_i$ for $i=1$ to $q$ as depicted
in \figref{fig:cliqueexample}.
The partition function at high $T$ or high $q$ with $T\gg T_\times$ is
\cite{ref:huCDPTlong}
\begin{equation}
  Z(G;q,\mathbf{v}) \approx q^N \prod_{i=1}^q \prod_{j=1}^{n_i} 
    \left(1 + \sum_{k=1}^{j-1} \frac{v_{k+\ell_j}}{q} \right)
  \label{eq:ZcliqueslargeqT}
\end{equation}
where $\ell_j = (j-1)(j-2)/2$.
\Eqnref{eq:ZcliqueslargeqT} is equivalent to \eqnref{eq:ZarbitraryGlargeqkG} 
to first order in $v_k$.
When $T\gg T_\times$, high $q$ results in the same approximation as high $T$ 
affirming the implication made with \eqnsref{eq:farbitraryGWsumT}{eq:farbitraryGWsumq}.

For high $T$ specifically, we make the additional approximation $v_k\approx \beta J_k$, 
and the free energy per site becomes 
\begin{equation}
  f \approx -k_B T\ln{} q - \frac{E_N}{q}
  \label{eq:fCliquesGWsumq}
\end{equation}
using the same approximations as in \eqnref{eq:farbitraryGWsumve}.
$E_N = \sum_{e'} J_{e'}/N$ is the energy per site.
In ref.\ \cite{ref:huCDPTlong}, we derived 
\begin{equation}
  T_\times^\mathrm{NIC}\approx \frac{(n-1) J}{2 k_B \ln{} q}
  \label{eq:triTcross}
\end{equation}
for constant $J$ and $q$ non-interacting cliques of \emph{fixed} 
size $n$. With $d=n-1$ edges per node, the result coincides well 
with \eqnsrefthree{eq:Tcrossoverd}{eq:TcrossoverMF}{eq:TcrossFE}.

\section{Thermal annealing comments} \label{sec:annealing}
For heat bath or simulated annealing (SA) algorithms in CD, when
$T\ll L|J_e|/[Nk_B\ln{} q]$, the global system remains in an ordered state.
For higher $T$, the \emph{global} system becomes increasingly disordered
in terms of partition function state probabilities.
If we consider states ``near'' equilibrium, small fluctuations in the energy
result in only tiny changes to the state probabilities through the Boltzmann
weight.
Another perspective is that even for a system of non-interacting cliques 
depicted in \figref{fig:cliqueexample}, larger $q$ creates a greater 
probability that a non-negligible fraction of cliques will be disconnected 
at a given $T$ at any given point during the stochastic solution.

Most SA algorithms (\eg{}, \cite{ref:smcd}) utilize energy \emph{differences}
to evaluate dynamic changes to the community division, so the system is 
effectively solved locally (algorithmically speaking, global parameters 
in the \emph{cost function} are a separate issue
\cite{ref:gn,ref:fortunato,ref:smcd,ref:kumpulaResLim,ref:rzlocal,
ref:lancfortunatomod,ref:traaglocalscope,ref:xiangmultireslimit}).
In practice, SA is limited to systems with $O(10^4)$ nodes without significant 
parallelization, so greedy algorithms ($T=0$) are used on the largest systems 
\cite{ref:blondel,ref:LPA,ref:rzlocal}.
SA implements a cooling scheme to fine tune solutions determined by the high
$T$ optimization, and our results indicate that cooling becomes more important
for large-$q$ systems.

\section{Conclusion} \label{sec:conclusion}
We showed a global disorder transition at a large number of communities
$q$ for a \emph{general} weighted (or unweighted) Potts model over essentially
\emph{arbitrary} graphs. 
The community structure of a complex network may be \emph{globally} disordered 
at large $q$ but still be \emph{locally} ordered and locally solvable.
Our results encompass many popular cost functions utilized for community detection,
including modularity and common Potts model variants.
We demonstrated this effect using stringent exact bounds as well as related results 
suggested by mean-field and other general approaches. 
With these bounds, results for a local system that occupies only a sub-volume 
of the original system lead to rigorous results for the full system, and they 
may have similar applications in the analysis of other hard computational 
problems where mean-field approaches are commonly applied. 
We also illustrated that in the strongest possible model partition, that 
of non-interacting cliques, the large-$q$ limit induces disorder akin to random 
thermal effects.

Increasing $q$ emulates increasing $T$ in arbitrary graphs for any CD method
that may be represented as a general weighted Potts model.
The asymptotic behavior of the global disorder transition varies slowly in $q$,
$T_\times\approx L|J_e|/[Nk_B\ln{} q]$, meaning that problems of practical
size maintain a finite region of solvability given a stochastic heat bath
algorithm.
Local algorithm dynamics (even for models which incorporate global weighting
parameters) serve to circumvent the global disorder transition.
This global disorder is generally circumvented by the often used SA algorithm,
but ``glassy'' problems with high noise (many extraneous intercommunity edges)
would remain a challenge for any algorithm or model.

\acknowledgments
This work was supported by NSF grant DMR-1106293 (ZN).
We also wish to thank M. Biskup, S. Chakrabarty, L. Chayes, V. Dobrosavljevic,
P. Johnson, and L. Zdeborov{\'a} for discussions and ongoing work.



\begin{thebibliography}{50}
\expandafter\ifx\csname url\endcsname\relax\def\url#1{\texttt{#1}}\fi

\bibitem{ref:newmanphystoday}
\Name{Newman M. E.~J.} \REVIEW{Phys. Today}{61}{2008}{33}.

\bibitem{ref:fortunatophysrep}
\Name{Fortunato S.} \REVIEW{Phys. Rep.}{486}{2010}{75}.

\bibitem{ref:clausetlarge}
\Name{Clauset A., Newman M. E.~J. \and Moore C.} \REVIEW{Phys. Rev.
  E}{70}{2004}{066111}.

\bibitem{ref:smcd}
\Name{Reichardt J. \and Bornholdt S.} \REVIEW{Phys. Rev. E}{74}{2006}{016110}.

\bibitem{ref:blondel}
\Name{Blondel V.~D., Guillaume J.-L., Lambiotte R. \and Lefebvre E.} \REVIEW{J.
  Stat. Mech.}{10}{2008}{P10008}.

\bibitem{ref:rosvallmaprw}
\Name{Rosvall M. \and Bergstrom C.~T.} \REVIEW{Proc. Natl. Acad. Sci.
  U.S.A.}{105}{2008}{1118}.

\bibitem{ref:gudkov}
\Name{Gudkov V., Montealegre V., Nussinov S. \and Nussinov Z.} \REVIEW{Phys.
  Rev. E}{78}{2008}{016113}.

\bibitem{ref:kumpulacliqueperc}
\Name{Kumpula J.~M., Kivel{\"a} M., Kaski K. \and Saram{\"a}ki J.}
  \REVIEW{Phys. Rev. E}{78}{2008}{026109}.

\bibitem{ref:lanc}
\Name{Lancichinetti A., Fortunato S. \and Kert{\'e}sz J.} \REVIEW{New J.
  Phys.}{11}{2009}{033015}.

\bibitem{ref:rzmultires}
\Name{Ronhovde P. \and Nussinov Z.} \REVIEW{Phys. Rev. E}{80}{2009}{016109}.

\bibitem{ref:barberLPA}
\Name{Barber M.~J. \and Clark J.~W.} \REVIEW{Phys. Rev. E}{80}{2009}{026129}.

\bibitem{ref:lancLFRcompare}
\Name{Lancichinetti A. \and Fortunato S.} \REVIEW{Phys. Rev.
  E}{80}{2009}{056117}.

\bibitem{ref:chengshen}
\Name{Cheng X.-Q. \and Shen H.-W.} \REVIEW{J. Stat. Mech.}{04}{2010}{P04024}.

\bibitem{ref:shenchengspectral}
\Name{Shen H.-W. \and Cheng X.-Q.} \REVIEW{J. Stat. Mech.}{10}{2010}{P10020}.

\bibitem{ref:rzlocal}
\Name{Ronhovde P. \and Nussinov Z.} \REVIEW{Phys. Rev. E}{81}{2010}{046114}.

\bibitem{ref:huCDPTsgd}
\Name{Hu D., Ronhovde P. \and Nussinov Z.} \REVIEW{Phil. Mag.}{92}{2012}{406}.

\bibitem{ref:goodMC}
\Name{Good B.~H., de~Montjoye Y.-A. \and Clauset A.} \REVIEW{Phys. Rev.
  E}{81}{2010}{046106}.

\bibitem{ref:nadakuditiSBM}
\Name{Nadakuditi R.~R. \and Newman M. E.~J.} \REVIEW{Phys. Rev.
  Lett.}{108}{2012}{188701}.

\bibitem{ref:reichardtstruct}
\Name{Reichardt J. \and Leone M.} \REVIEW{Phys. Rev. Lett.}{101}{2008}{078701}.

\bibitem{ref:decelleKMZPT}
\Name{Decelle A., Krzakala F., Moore C. \and Zdeborov{\' a} L.} \REVIEW{Phys.
  Rev. Lett.}{107}{2011}{065701}.

\bibitem{ref:dorogovtsevRMP}
\Name{Dorogovtsev S.~N., Goltsev A.~V. \and Mendes J. F.~F.} \REVIEW{Rev. Mod.
  Phys.}{80}{2008}{1275}.

\bibitem{ref:bashanphystrans}
\Name{Bashan A., Bartsch R.~P., Kantelhardt J.~W., Havlin S. \and Ivanov P.~C.}
  \REVIEW{Nature Comm.}{3}{2012}{702}.

\bibitem{ref:HRNimages}
\Name{Hu D., Ronhovde P. \and Nussinov Z.} \REVIEW{Phys. Rev.
  E}{85}{2012}{016101}.

\bibitem{ref:juhaszrbPMlargeq}
\Name{Juh\'asz R., Rieger H. \and Igl\'oi F.} \REVIEW{Phys. Rev.
  E}{64}{2001}{056122}.

\bibitem{ref:mercaldorbPMlargeq}
\Name{Mercaldo M.~T., d'Auriac J.-C.~A. \and Igl{\' o}i F.} \REVIEW{Europhys.
  Lett.}{70}{2005}{733}.

\bibitem{ref:changlargeqzeroes}
\Name{Chang S.-C. \and Shrock R.} \REVIEW{Int. J. Mod. Phys. B}{21}{2007}{979}.

\bibitem{ref:arenasmultires}
\Name{Arenas A., Fern{\'a}ndez A. \and G{\'o}mez S.} \REVIEW{New J.
  Phys.}{10}{2008}{053039}.

\bibitem{ref:kumpulamultires}
\Name{Kumpula J.~M., Saram{\"a}ki J., Kaski K. \and Kert{\'e}sz J.}
  \REVIEW{Fluct. Noise Lett.}{7}{2007}{L209}.

\bibitem{ref:rosvallmultires}
\Name{Rosvall M. \and Bergstrom C.~T.} \REVIEW{PLoS ONE}{6}{2011}{e18209}.

\bibitem{ref:leskoveclargenetworks}
\Name{Leskovec J., Lang K.~J., Dasgupta A. \and Mahoney M.~W.} \REVIEW{Internet
  Mathematics}{6}{2009}{29}.

\bibitem{ref:gn}
\Name{Newman M. E.~J. \and Girvan M.} \REVIEW{Phys. Rev. E}{69}{2004}{026113}.

\bibitem{ref:traagPRE}
\Name{Traag V.~A. \and Bruggeman J.} \REVIEW{Phys. Rev. E}{80}{2009}{036115}.

\bibitem{ref:reichardt}
\Name{Reichardt J. \and Bornholdt S.} \REVIEW{Phys. Rev.
  Lett.}{93}{2004}{218701}.

\bibitem{ref:traaglocalscope}
\Name{Traag V.~A., Van~Dooren P. \and Nesterov Y.} \REVIEW{Phys. Rev.
  E}{84}{2011}{016114}.

\bibitem{ref:LPA}
\Name{Raghavan U.~N., Albert R. \and Kumara S.} \REVIEW{Phys. Rev.
  E}{76}{2007}{036106}.

\bibitem{ref:blatt}
\Name{Blatt M., Wiseman S. \and Domany E.} \REVIEW{Phys. Rev.
  Lett.}{76}{1996}{3251}.

\bibitem{ref:ispolatov}
\Name{Ispolatov I., Mazo I. \and Yuryev A.} \REVIEW{J. Stat. Mech.}{09}{2006}{P09014}.

\bibitem{ref:hastings}
\Name{Hastings M.~B.} \REVIEW{Phys. Rev. E}{74}{2006}{035102}.

\bibitem{ref:ellismonaghanexternalH}
\Name{Ellis-Monaghan J.~A. \and Moffatt I.} \REVIEW{Adv. App.
  Math.}{47}{2011}{772 }.

\bibitem{ref:fortunato}
\Name{Fortunato S. \and Barth{\'e}lemy M.} \REVIEW{Proc. Natl. Aca. Sci.
  U.S.A.}{104}{2007}{36}.

\bibitem{ref:kumpulaResLim}
\Name{Kumpula J.~M., Saram{\"a}ki J., Kaski K. \and Kert{\'e}sz J.}
  \REVIEW{Euro. Phys. J. B}{56}{2007}{41}.

\bibitem{ref:lancfortunatomod}
\Name{Lancichinetti A. \and Fortunato S.} \REVIEW{Phys. Rev.
  E}{84}{2011}{066122}.

\bibitem{ref:huCDPTlong}
\Name{Hu D., Ronhovde P. \and Nussinov Z.} \REVIEW{e-print
  arXiv:1204.4167}{}{2012}{}.

\bibitem{ref:batista_nussinov_Dreduc}
\Name{Batista C.~D. \and Nussinov Z.} \REVIEW{Phys. Rev. B}{72}{2005}{045137}.

\bibitem{ref:nussinovholographies}
\Name{Nussinov Z., Ortiz G. \and Cobanera E.} \REVIEW{e-print
  arXiv:1110.2179}{}{2011}{}.

\bibitem{ref:jacksonmvtutte}
\Name{Jackson B. \and Sokal A.~D.} \REVIEW{J. Combinatorial Theory, Series
  B}{99}{2009}{869}.

\bibitem{ref:pottslargeqmean}
\Name{Mittag L. \and Stephen M.~J.} \REVIEW{J. Phys. A: Math. Nucl.
  Gen.}{7}{1974}{L109}.

\bibitem{ref:biskuppotts}
\Name{Biskup M., Chayes L. \and Crawford N.} \REVIEW{J. Stat.
  Phys.}{122}{2006}{1139}.

\bibitem{ref:pearcelargeqmean}
\Name{Pearce P.~A. \and Griffiths R.~B.} \REVIEW{J. Phys. A: Math.
  Gen.}{13}{1980}{2143}.

\bibitem{ref:xiangmultireslimit}
\Name{Xiang J. \and Hu K.} \REVIEW{Physica A}{391}{2012}{4995}.

\end{thebibliography}
\end{document}